\documentclass[
  prd,
  onecolumn,
  superscriptaddress,
  nofootinbib,
  amsmath,
  amssymb,
  aps,
]{revtex4-2}

\pdfoutput=1

\usepackage[utf8]{inputenc}
\usepackage{graphicx}
\usepackage{bm}
\usepackage{microtype}
\usepackage{threeparttable}
\usepackage{mathtools}
\usepackage[colorlinks=true]{hyperref}

\usepackage{cleveref}
\crefname{figure}{Fig.}{Fig.}
\crefname{equation}{Eq.}{Eq.}
\crefname{equation}{Eq.}{Eqs.}

\newcommand{\p}{\partial}

\newcommand{\f}[2]{\frac{#1}{#2}}

\newcommand{\be}{\begin{equation}}
\newcommand{\ee}{\end{equation}}

\newcommand{\ed}{{\cal E}}       
\newcommand{\pa}{{\cal A}}

\begin{document}

\author{Michał Spaliński}
\email{michal.spalinski@ncbj.gov.pl}
\affiliation{National Centre for Nuclear Research, 02-093 Warsaw, Poland}
\affiliation{Physics Department, University of Białystok,
15-245 Bia\l ystok, Poland}

\title{ Far from equilibrium attractors in phase space }

\begin{abstract}

The emergence of far from equilibrium, prehydrodynamic attractors is an
important feature of boost-invariant flow in models of relativistic fluid
dynamics, as well as in some microscopic theories. Originally, these attractors
were defined in terms of attractor solutions, using a partial decoupling of the
equations of motion that relied on using special variables. Reliance on such a
decoupling restricts the class of systems that can be analysed.  Instead of
introducing special variables, here we directly leverage the singularity of the
evolution equations at early proper time. This singularity is a consequence of
boost invariance, which should be regarded as crucial physical input stemming
from fundamental properties of particle production in QCD. We posit that it
provides initial conditions which determine the attractor hypersurface in phase
space, irrespective of whether the evolution equations can be partially
decoupled or not.  We validate this in a case where the equations of motion
cannot be decoupled but the attractor can still be identified and governs the
behaviour of generic solutions in a similar way to what happens in cases where
attractor solutions exist.

\end{abstract}

\maketitle

\section{Introduction}
\label{sec:intro}

At the macroscopic level, the evolution of complex systems toward equilibrium
involves an effective loss of memory of the initial conditions. While the
approach to equilibrium is captured by the hydrodynamic gradient expansion at
sufficiently late times, it can happen that universal behaviour sets in
significantly earlier, well before it can be attributed to hydrodynamics. Such a
situation may be interpreted as a prehydrodynamic attractor.  An important
example appears in the context of quark-gluon plasma (QGP) physics, where
Mueller-Israel-Stewart theory~\cite{Muller:1967zza,Israel:1976tn,Israel:1979wp}
(MIS) is a key element of the phenomenological description of heavy ion
collisions. Simulations of QGP dynamics involve flows
such that at the early stages the energy-momentum tensor is very
anisotropic~\cite{Schenke:2021mxx}. The success of these simulations in describing experimental data
suggests a measure of universality emerging while the system is still very far
from equilibrium. Indeed, an idealised description of QGP dynamics by Bjorken
flow within MIS theory features an early-time, prehydrodynamic
attractor~\cite{Heller:2015dha}. A similar situation arises for Bjorken flow
within kinetic theory, where the early-time regime is characterised by
free-streaming while the emergence of hydrodynamic behaviour at late times
signals the dominance of interactions~\cite{Blaizot:2017ucy,Blaizot:2019scw}.

A key feature of these models is that the equations of motion are singular at
early time.  This singularity can be traced to the dominance of the longitudinal
expansion in the initial stages of a heavy-ion collision. Its origins lie in the
set of approximations underlying Bjorken flow: 
longitudinal boost invariance and homogeneity in the transverse plane. Of these,
the crucial dynamical assumption is boost invariance, which
is a characteristic property of particle production in QCD at high
energies~\cite{Bjorken:1982qr}. Imposing boost invariance on MIS theory introduces dynamical
information that is essential for the emergence of early-time attraction --- this
may explain why MIS theory is able to approximate the significantly more complex
dynamics of QCD. Boost invariance is also a key element of the emergence of
hydrodynamic behaviour in other studies of early-time
physics~\cite{Carrington:2020ssh,Carrington:2021qvi,Carrington:2024utf}.  The
importance of boost invariance in connection with early-time attraction was
highlighted in the recent paper~\cite{Chen:2024pez} (see also
Ref.~\cite{Nugara:2023eku}), where it was found that breaking boost-invariance in
the initial conditions tends to wash out the early-time attraction, in contrast
to relaxing the condition of transverse
homogeneity~\cite{Kurkela:2019set,Ambrus:2021sjg,Ambrus:2021fej,Ambrus:2022koq,Nugara:2024net,An:2023yfq}.

The appearance of the attractor solution in conformal MIS theory relies on a
partial decoupling of the equations of motion, which can be achieved be a
suitable change of variables; specifically, by introducing a dimensionless
``clock variable'' $w\equiv\tau T$, where $\tau$ is the proper time and $T$ is
the effective temperature. The attractor is then described by a specific
solution of a single ordinary differential equation which determines the
pressure anisotropy $\pa$ as a function of $w$. However, in less constrained
scenarios such a decoupling is typically not possible, and this poses
a technical obstacle to the exploration of attractors in more general settings.

A natural re-framing of the problem, which applies regardless of whether a
decoupling of the equations of motion can be brought about, is to recognise
attractors as submanifolds of phase space where generic solutions
congregate. This perspective is general enough to accommodate many types of
attractor phenomena. In this context, Ref.~\cite{Heller:2020anv} introduced a
local notion of attractor that avoids referencing any asymptotic limits --- as
dimensionality reduction of phase space regions. It was shown
there that for conformal MIS theory generic solutions converge to a region of
phase space coinciding with a hypersurface determined from the known attractor
solution.  This raised the question whether the attractor hypersurface could be
found directly without appealing to an attractor solution in the sense of
Ref.~\cite{Heller:2015dha}. Recently, this question was resolved in
Ref.~\cite{An:2023yfq}, where it was shown that the attractor hypersurface of
conformal MIS theory in phase space can be found directly using a regularity
condition at asymptotically small proper times.

The remaining piece of the puzzle is to clarify the relationship between the
attractor hypersurface determined in this way and the concept of an attractor
solution in the sense of Ref.~\cite{Heller:2015dha}. We will resolve this issue
by showing that these notions are identical in cases when an attractor solution
is available. In brief, an explicit description of the attractor hypersurface
can be obtained by means of its intersections with constant proper time
slices. Generally, these sections evolve from slice to slice; however, in the
special cases where the equations can be partially decoupled, they can be
projected and the resulting curve coincides with the attractor solution.  This
is illustrated below, in \cref{sec:confomis,sec:dn}, by considering two models
of MIS theory where this projection comes about in two rather different ways.

These developments lead to a broader question: can the singularity at early
times be leveraged to determine the attractor hypersurface also in cases when
the equations of motion cannot be partially decoupled?  We suggest that the
answer is affirmative. We provide evidence in support of this assertion in
\cref{sec:nonconf}, where we analyse a model derived from conformal MIS theory
via a conformal symmetry-breaking modification of the equation of
state. Although this model cannot be effectively treated using the original
approach of Ref.~\cite{Heller:2015dha}, the regularity condition at early proper
time unambiguously determines the attractor which captures the collective
behaviour of generic solutions in a manner qualitatively similar to what is seen
in conformal MIS theory.

We also point out that in all the models discussed here, an analytic description
of the attractor sections on constant proper time slices can be obtained in
parametric form.  This comes about by reinterpreting the transseries solutions
in proper time as functions of the initial data. In particular, in examples
where an attractor solution $\pa(w)$ exists, one can formally recover its
transseries form from the transseries expansion in proper time.

\section{Conformal MIS theory}
\label{sec:confomis}

In this section we describe how the attractor hypersurface of conformal MIS
theory is determined, with an emphasis on the role of early time asymptotics. We
will also show how this surface can be projected when the special coordinates
$\pa, w$ are used to parameterise the phase space, thus resolving the
question concerning the relationship of the attractor solution to the attractor
hypersurface.

MIS theory is expressed in terms of the classical fields $\ed$ (the energy
density), $u^\mu$ (the flow velocity) and $\pi^{\mu\nu}$ (the shear-stress
tensor). They satisfy the following set of partial
differential equations
\begin{subequations}
  \begin{align}
    u\cdot\nabla\ed&=-(\ed+p)\nabla\cdot u+ u^\nu\nabla^\mu\pi_{\mu\nu},
    \label{eq:MIS.a}
    \\
    (\ed+p)u\cdot\nabla u_\mu&=-\Delta_{\mu\nu}\nabla^\nu p-\Delta_{\mu\nu}\nabla_\lambda\pi^{\nu\lambda},
    \label{eq:MIS.b}
    \\
    \Delta_{\mu\alpha}\Delta_{\nu\beta}u\cdot\nabla\pi^{\alpha\beta}&=-\left(1+\frac{4}{3}\tau_\pi\nabla\cdot u\right)\pi_{\mu\nu}-2\eta\sigma_{\mu\nu},
    \label{eq:MIS.c}
  \end{align}
\end{subequations}
where $\nabla_\mu$ is the covariant derivative, $\Delta_{\mu\nu}\equiv
g_{\mu\nu}+u_\mu u_\nu$ is the transverse projector,
$\sigma_{\mu\nu}=\frac{1}{2}\Delta_{\mu\alpha}\Delta_{\nu\beta}(\nabla^\alpha
u^\beta+\nabla^\beta u^\alpha-\frac{2}{3}\Delta^{\alpha\beta}\nabla\cdot u)$ is
the shear tensor, $\eta$ is the shear viscosity and $\tau_\pi$ is the
relaxation time. The first two of the above equations express conservation of
the energy-momentum tensor, while the last is the MIS equation which determines
the evolution of the shear-stress tensor. Throughout this paper we set the bulk
viscosity as well as the bulk pressure to zero.
In this section we will also adopt equations of state and transport coefficients
dictated by conformal invariance:
\be
\label{eq:conformal}
\ed=3p=a\, T^4,
\qquad
\eta=\frac{4}{3} a\, C_\eta T^3,
\qquad \tau_\pi=C_\pi/T,
\ee
where $C_\eta$, $C_{\pi}$ and $a$ are dimensionless constants and $T$ is the
effective temperature.

With the above assumptions, denoting lab frame time by $t$ and the coordinate
along the collision axis by $z$,
the energy-momentum tensor for
Bjorken flow can be parameterised in terms of two functions of the proper time
$\tau\equiv \sqrt{t^2-z^2}$: the effective temperature $T$, and the pressure
anisotropy $\pa\equiv 9\pi^k_k/2\ed$. The physical meaning of the pressure
anisotropy is manifest if one expresses it in terms of eigenvalues of the
energy-momentum tensor $(T_\mu^\nu) \equiv \mathrm{diag} ( -\ed, p_L, p_T, p_T)$
as $\pa = (p_T - p_L)/p$, (for details see,
e.g.~\cite{Spalinski:2022cgj,Jankowski:2023fdz}).
The equations of
motion given in \cref{eq:MIS.a,eq:MIS.b,eq:MIS.c} can then be written in a form describing
evolution in proper time as~\cite{An:2023yfq}
\begin{subequations}
  \begin{align}
    \tau\partial_\tau\ln T &=-\f{1}{3}+\f{1}{18}\pa,
    \label{eq:misbjT}\\
    \tau\partial_\tau\pa &=
    \frac{8 C_\eta}{C_\pi } - \frac{\tau T}{C_\pi} \pa- \f{2}{9} \pa^2.
    \label{eq:misbjA}
  \end{align}
\end{subequations}
This system of equations is not autonomous: in addition to $T, \pa$ one also
needs the value of $\tau$ to fully describe the state at a given
instant. Formally, such non-autonomous systems can be equivalently presented in
an autonomous form at the price of introducing an additional variable (see
e.g. ~\cite{Wiggins}). The resulting extended phase space is
thus effectively $3$-dimensional, parameterised by $(T, \pa, \tau)$.  In what
follows, it will be most convenient to think of this extended phase space as a
set slices at constant values of the proper time. On each such slice at some
specified value of $\tau$, a solution is represented by the point $(T(\tau),
\pa(\tau))$.

\begin{figure}[t]
  \centering
  \includegraphics[width=0.45\linewidth]{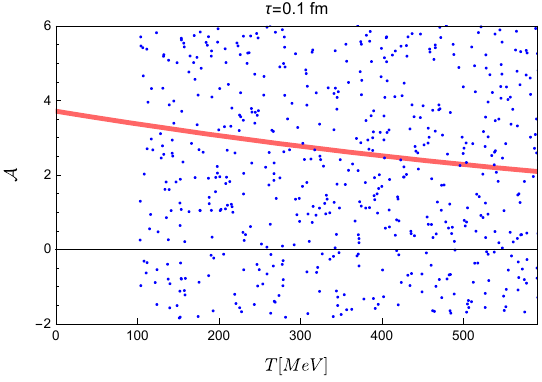}
  \includegraphics[width=0.45\linewidth]{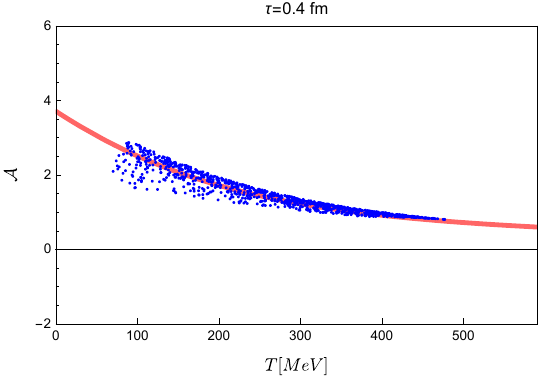}
  \includegraphics[width=0.45\linewidth]{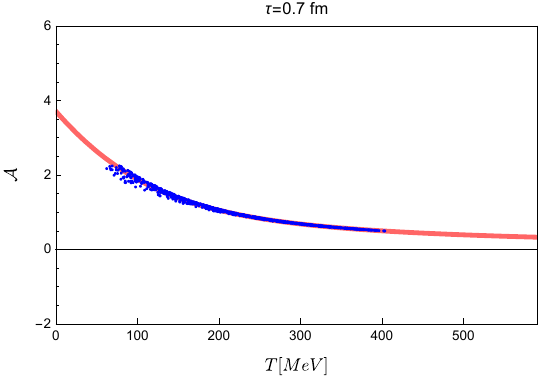}
  \includegraphics[width=0.45\linewidth]{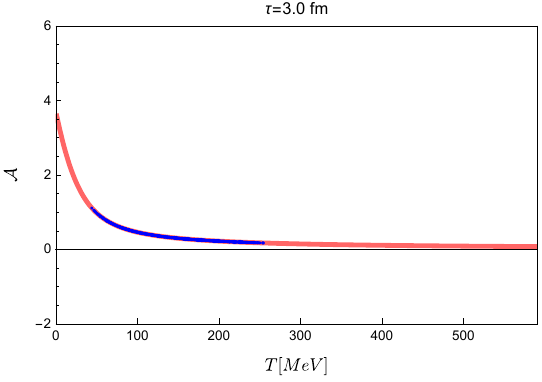}
  \caption{
    The evolution of the attractor section (the red curve) on four constant proper
    time slices in the extended phase space of conformal MIS theory parameterised by $(T, \pa, \tau)$.  At
    the initial time (taken as $0.1$ fm) the initial conditions for the solutions
    plotted as blue dots are set uniformly in a range of initial temperatures and
    pressure anisotropies. At $\tau=0.7$ fm all the solutions are in the vicinity
    of the attractor, with pressure anisotropies ranging from around $0.5$ to
    about $2.5$.  Subsequent evolution follows the (evolving) attractor. The range of
    temperatures and pressure anisotropies becomes narrower as time goes on,
    tending toward zero as the system cools.
  }
  \label{fig:attra}
\end{figure}

A crucial fact is that the system of evolution equations given in
\cref{eq:misbjT,eq:misbjA} is singular at $\tau=0$, so solutions are defined only on the
domain $\tau>0$. There are however two special families of solutions (each labelled
by a single integration constant denoted below by $\mu$) for which the pressure
anisotropy is regular at $\tau=0$.  Denoting asymptotic equality by $\sim$, one
finds that as $\tau\to0$
\begin{equation}
  \pa \sim\pa_\pm\,,
  \qquad
  \qquad
  T \sim (\mu\tau)^{\beta-1}\,,
  \label{eq:attractorA}
\end{equation}
where 
\begin{equation}
  \label{eq:attra}
  \pa_\pm \equiv \pm 6 \sqrt{\f{C_\eta}{C_\pi}}
\end{equation}
and
\begin{equation}
  \beta=\frac{\sqrt{C_\eta }+2 \sqrt{C_\pi }}{3 \sqrt{C_\pi}}\,.
\end{equation}
The family of solutions with $\pa\sim\pa_+$ as $\tau\to 0$
defines the attractor hypersurface in phase space. These initial conditions can
be used to calculate series expansions that provide an analytic representation
of the attractor at early time~\cite{An:2023yfq}. Since the extended phase
space is $3$-dimensional here, sections of the attractor are curves in the $(T,
\pa)$ planes at constant values of $\tau$. The behaviour of generic solutions is
shown in \cref{fig:attra}.

One can also describe the attractor analytically
in the hydrodynamic regime, where approximate asymptotic
solutions are provided by the late proper time expansion (related to
the hydrodynamic gradient expansion, see, e.g.,
Refs.~\cite{Florkowski:2017olj,Soloviev:2021lhs,Jankowski:2023fdz}):
\cref{eq:misbjT,eq:misbjA} imply the well-known asymptotic behaviour of generic solutions
%
\begin{equation}
  \label{eq:bjorken}
  T \sim \Lambda (\Lambda\tau)^{-\frac{1}{3}} \,,
  \qquad\quad
  \qquad\quad
  \pa \sim 8C_\eta(\Lambda\tau)^{-\frac{2}{3}}\,,
\end{equation}
where $\Lambda$ is an integration constant which depends on the initial
conditions. One usually views these relations as expressing the proper time
dependence of a specific solution $(T(\tau),\pa(\tau))$ with some value of
$\Lambda$. Here we propose a change of perspective: to interpret these equations as a
parametric representation of a curve $(T(\Lambda),\pa(\Lambda))$ on the slice of
phase space at a fixed value of $\tau$.  This implies that on a proper time
slice at a sufficiently large value of the proper time, generic solutions lie
asymptotically close to this curve, which is a section of the hydrodynamic
attractor on that particular time slice. Obviously, the shape of this curve
changes from slice to slice. The value of $\Lambda$ determines which point on
the attractor section corresponds to a specific solution (or, equivalently, a
specific initial condition).

By including subleading terms, along with the correct transseries
contributions~\cite{An:2023yfq}, one can extend the validity of \cref{eq:bjorken}
to early times. Thus, at least in principle, sections of the attractor can be
parametrically determined on any time slice.  However, this is not a practical
method at early times, because it is not known a priori which value of the
transseries parameter singles out the attractor~\cite{Heller:2015dha,
Aniceto:2022dnm}.  In practice, one can determine the attractor locus
numerically by imposing the initial condition $\pa(\tau_0)=\pa_+$ along with
initial conditions for $T(\tau_0)$ spanning a range of temperatures at some very
small proper time $\tau_0>0$. In this way, a set of solutions is obtained which
threads the attractor surface in phase space. One can then determine the
attractor section on any slice by evaluating these solutions on that slice and
interpolating -- this is how the red curves in \cref{fig:attra} were
determined. It is clear that the attractor sections evolve from slice to
slice. \cref{fig:attra} also demonstrates the three distinct stages of
evolution noted in~\cite{Heller:2020anv} (see also Ref.~\cite{An:2023yfq}).
The
first stage is a rapid approach to the nonequilibrium part of the attractor
which can be attributed to the longitudinal expansion; the second stage
describes descent onto the attractor and is characterised by the decay of
exponentially damped nonhydrodynamic degrees of freedom; and then the third,
hydrodynamic stage, encompassing evolution along the attractor which is captured
by the gradient expansion.  It is also clear from \cref{fig:attra} that on a
given proper time slice solutions with higher temperature are closer to the
attractor than solutions with smaller temperatures.

\begin{figure}[t]
  \centering
  \includegraphics[width=0.45\linewidth]{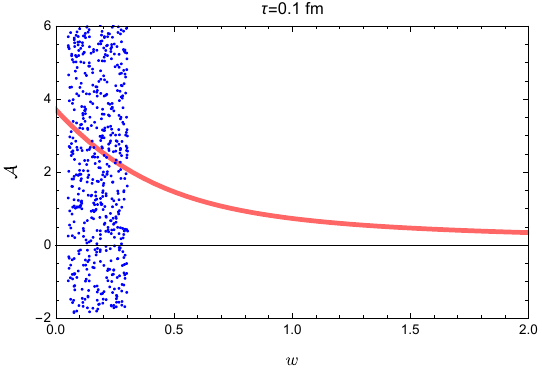}
  \includegraphics[width=0.45\linewidth]{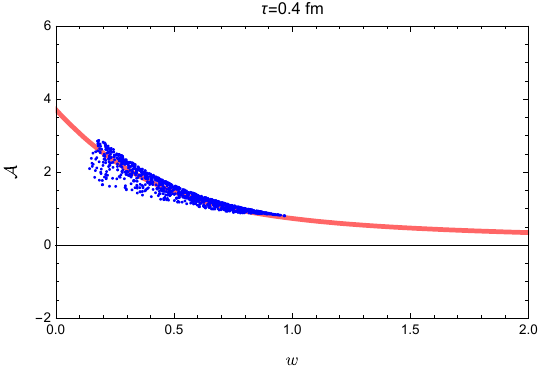}
  \includegraphics[width=0.45\linewidth]{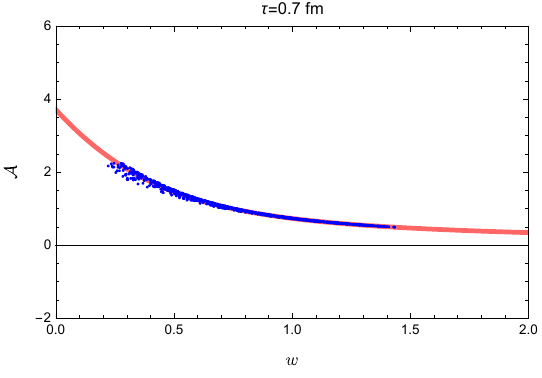}
  \includegraphics[width=0.45\linewidth]{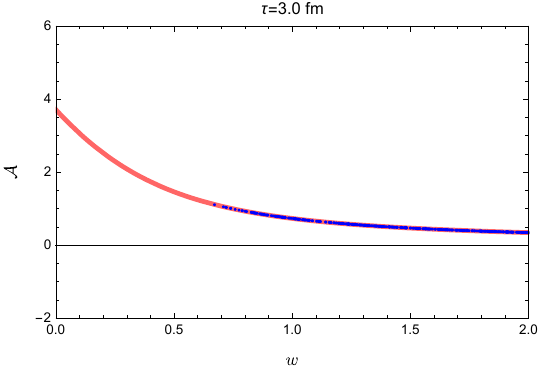}
  \caption{
    The attractor section (the red curve) on the same set of four constant
    proper time slices as in \cref{fig:attra}, but now with the
    extended phase space parameterised by $(w, \pa, \tau)$. Note that in this
    parameterisation the attractor sections are identical on all slices. The blue
    dots represent exactly the same solutions as in \cref{fig:attra}. By
    $\tau=0.7$ fm all solutions are in the vicinity of the attractor, with
    pressure anisotropies ranging from around $0.5$ to about $3.5$. Subsequent evolution
    follows the attractor; the range of values of $w$ becomes narrower and drifts to
    higher values as time goes on. The range of pressure anisotropies of course also
    narrows, tending toward zero.
  }
  \label{fig:attraw}
\end{figure}

We now turn to an important issue raised implicitly by Ref.~\cite{An:2023yfq}:
what is the precise connection between the attractor hypersurface in phase space
and the attractor solution in the sense of Ref.~\cite{Heller:2015dha}?  We will
show that these two notions coincide, and they do so at all times. To make this
connection, we will now use the dimensionless variable $w\equiv\tau T$ as one of
the extended phase space coordinates instead of $T$. With this choice, the slice
at each value of $\tau$ is parameterised by $(w, \pa)$.
It is simplest to begin with the hydrodynamic regime, where \cref{eq:bjorken}
can be reinterpreted as representing the attractor section on a constant~$\tau$
slice in the $(w, \pa)$ variables:
\begin{equation}
  \label{eq:hydroslice}
  w(\Lambda) \sim (\Lambda\tau)^{\frac{2}{3}} \,,
  \qquad\quad
  \pa(\Lambda) \sim 8C_\eta(\Lambda\tau)^{-\frac{2}{3}}\,.
\end{equation}
These relations define a curve on this slice, expressed in parametric form, with $\Lambda$
playing the role of a parameter along it, labelling different solutions in this
regime. By eliminating the parameter $\Lambda$, one recognises this curve as the
well-known Navier-Stokes result $\pa\sim 8C_\eta/w$ (see
e.g. Ref.~\cite{Florkowski:2017olj}). In conformal models $\Lambda$ always
enters in the combination $\Lambda\tau$, so when it is eliminated the resulting
curve does not depend on $\tau$: it is the same across all slices in the late
proper time regime where \cref{eq:hydroslice} is valid.

The above observation can easily be extended to all times, because it remains true
also when the full transseries representation of the solution is used instead of
just the leading terms given in \cref{eq:hydroslice}:
\begin{subequations}
  \begin{align}
    w(\Lambda) &\sim (\Lambda\tau)^{2/3}
    \Phi_0(\Lambda\tau) +
    \sigma (\Lambda\tau)^{\frac{2 C_\eta}{3 C_\tau}} e^{-\frac{3}{2C_\tau}(\Lambda\tau)^{\frac{2}{3}}}
                 \Phi_1(\Lambda\tau) + \cdots
  \label{eq:mists.a}
    \\
    \mathcal{A}(\Lambda) &\sim
    \frac{8C_\eta}{(\Lambda\tau)^{2/3}}
    \Psi_0(\Lambda\tau) +
    \tilde{\sigma}(\Lambda\tau)^{\frac{2 C_\eta}{3 C_\tau}}
    e^{-\frac{3}{2C_\tau}(\Lambda\tau)^{\frac{2}{3}}}
    \Psi_1(\Lambda\tau) + \cdots
  \label{eq:mists.b}
  \end{align}
\end{subequations}
where $\Phi_k(\Lambda\tau), \Psi_k(\Lambda\tau)$ are power series in
$\Lambda\tau$, with $\Lambda$-independent coefficients determined by the
equations of motion, $\sigma$ is the transseries parameter and $\tilde{\sigma}$ is
related to it. Formally, eliminating the parameter $\Lambda$ order by order
leads to a curve which is independent of $\tau$. This means that when the phase
space is parameterised using $(w, \pa, \tau)$, the attractor section is {\em
identical across all proper time slices}. This conclusion can be corroborated
numerically, and is illustrated in \cref{fig:attraw}, which plots exactly the
same data as \cref{fig:attra}, but using the $(w, \pa, \tau)$ parameterisation
in place of $(T, \pa, \tau)$.

Since sections of the attractor hypersurface are the same on each slice, they
may be projected to the $(w, \pa)$ plane.  A hypersurface with this property is
sometimes called a cylindrical surface (or a generalised cylinder, see
e.g. Ref.~\cite{zwillinger}).  This projected curve can be
interpreted as a functional dependence $\pa(w)$ and coincides with the attractor
solution in the sense of Ref.\cite{Heller:2015dha}. Thus, one can in principle
recover the attractor solution $\pa(w)$ analytically --- as a transseries --- by
eliminating the parameter $\Lambda$ order by order in the transseries expansion
given in \cref{eq:mists.a,eq:mists.b}.

It is clear that this projection comes about due to the
fact that the equations of motion in the $(w, \pa, \tau)$ variables 
can be partially decoupled, in the sense that $\pa(w)$ satisfies~\cite{Heller:2015dha,Florkowski:2017olj}
\begin{equation}
  \label{eq:Aw}
C_\pi w \left(1 + \f{\pa}{12}\right) \p_w\pa = 12 C_\eta - \frac{3}{2} w \pa-
\f{1}{3} C_\pi \pa^2\,,
\end{equation}
while $w(\tau)$ can subsequently be determined from
\begin{equation}
  \label{}
     \tau\partial_\tau w =\f{2}{3} w \left(1 +\f{1}{12}\pa \right)\,.
\end{equation}
This is convenient, but from the phase space perspective it does not offer a
significant advantage, since the attractor hypersurface can also be identified
in other parameterisations.




\section{The Denicol-Noronha model}
\label{sec:dn}

In this section we briefly discuss another case when the evolution equations
partially decouple: this is the model introduced by Denicol and Noronha, who
found its attractor solution analytically~\cite{Denicol:2017lxn}~\footnote{I am
grateful to Gabriel Denicol for emphasising the significance of this example.}.
This model is defined by the MIS equations given in \cref{eq:MIS.a,eq:MIS.b,eq:MIS.c}, with the
same equations of state as in \cref{eq:conformal}, but with the transport
coefficients taken as
\be
\label{eq:dntcs}
\eta/s = C_\eta T/m\,,
\qquad \tau_\pi=C_\pi/m\,.
\ee
Here $m$ is a mass scale and $C_\eta$ and $C_\pi$ are dimensionless
constants.

In this model, the MIS equations of motion for Bjorken flow,
\cref{eq:misbjT,eq:misbjA}, take the form
\begin{subequations}
  \begin{align}
    \tau\partial_\tau\ln T &=-\f{1}{3}+\f{1}{18}\pa\,,
    \label{eq:dnT}\\
    \tau \p_\tau \pa &= \f{8 C_\eta}{C_\tau} - \f{\tau m}{C_\tau} \pa - \frac{2}{9} \pa^2\,.
    \label{eq:dnA}
  \end{align}
\end{subequations}
The MIS equation, \cref{eq:dnA}, is decoupled from the conservation equation,
\cref{eq:dnT}, without the need to introduce a new independent variable (such as
$w$ in conformal MIS theory). This example is
especially instructive, since the general solution of this equation is known.
However, for the present discussion, the essential point is that the equations of motion,
\cref{eq:dnT,eq:dnA}, are singular at $\tau=0$.  While solutions exist only for
$\tau>0$, there are two special solutions of \cref{eq:dnA} for which the
pressure anisotropy is regular at $\tau=0$, with the early time behaviour
$\pa\sim\pa_\pm$, where $\pa_\pm$ is given in \cref{eq:attra}.  The solution
with the upper sign acts as an attractor.

\begin{figure}[t]
  \centering
  \includegraphics[width=0.45\linewidth]{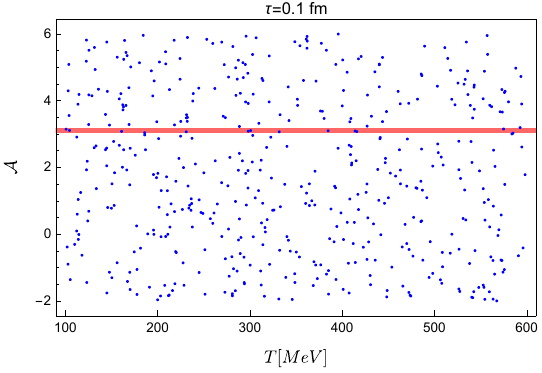}
  \includegraphics[width=0.45\linewidth]{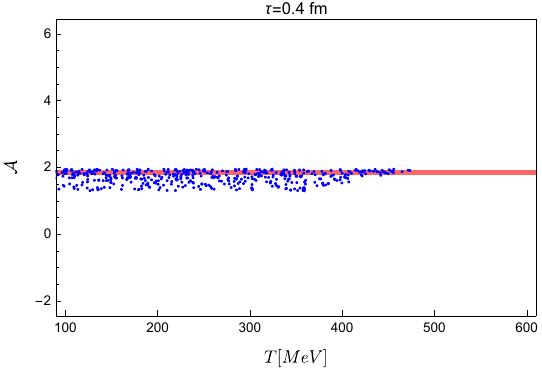}
  \includegraphics[width=0.45\linewidth]{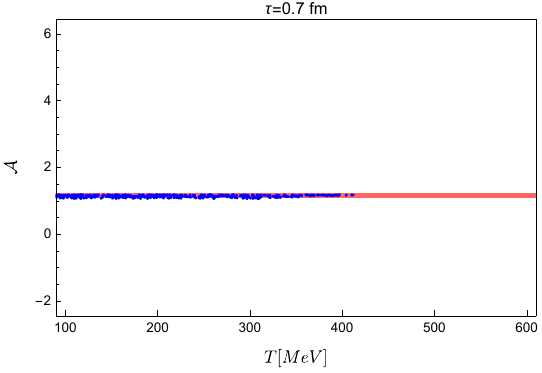}
  \includegraphics[width=0.45\linewidth]{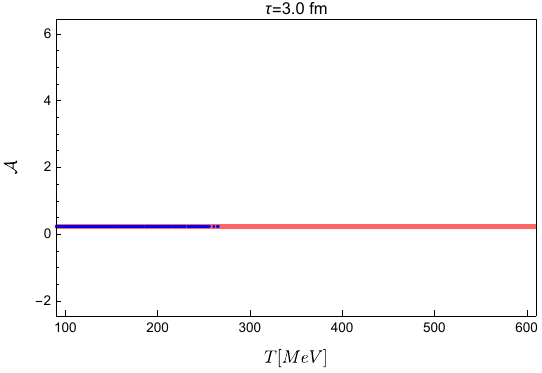}
  \caption{The evolution of the attractor section (the red curve) on a set of constant
    proper time slices in the extended phase space of the
    Denicol-Noronha model. At the initial time ($0.1$ fm)
    the initial conditions for the solutions plotted as blue dots are set uniformly
    in a range of initial temperatures and pressure anisotropies. By $\tau=0.7$ fm
    all the solutions are in the vicinity of the attractor, with pressure
    anisotropies of order $1$. Subsequent evolution follows
    the (evolving) attractor. The range of temperatures
    becomes narrower as time goes on, tending toward zero as the system cools.
  }
  \label{fig:crtattr}
\end{figure}

Here we reinterpret this attractor from the phase space perspective. Since the
equations of motion \cref{eq:dnT,eq:dnA} are non-autonomous, the phase space is
extended and parameterised by $(T, \pa, \tau)$.  The attractor can be described
in full analogy with the discussion in \cref{sec:confomis}: one can determine a
set of solutions threading the attractor hypersurface by solving the equations
of motion numerically, setting initial conditions consistent with the condition
of regularity of the pressure anisotropy at $\tau=0$. The results of this are
displayed in \cref{fig:crtattr}, which shows sections of the attractor along
with a set of generic solutions on four slices at fixed proper times.  The
attractor is evolving from slice to slice, but since it is constant in $T$, it
can be projected to the $(\tau, \pa)$ plane.  The projection coincides with the
attractor solution of \cref{eq:dnA}.

\section{A model of nonconformal MIS theory}
\label{sec:nonconf}


The previous sections have demonstrated that in two models where
the equations of motion can be partially decoupled one can determine the full
attractor hypersurface in phase space without appealing to the attractor
solution. The only essential feature of the dynamics
which had led to early time attraction was the singularity at early time. This
suggests that this may be a more general feature: that as long as the evolution
equations are singular at early times, the attractor hypersurface in phase space
can be determined using regularity conditions at $\tau=0$. This section provides
some evidence for this claim: we will consider a model of MIS theory where the
equations of motion can no longer be
partially decoupled and thus one cannot find an attractor solution in the sense
of Ref.~\cite{Heller:2015dha}. However, since the evolution equations are
singular at $\tau=0$, the final outcome is essentially the same: there is an
attractor hypersurface in phase space where generic solutions congregate in the
course of evolution.


We consider MIS theory with an equation of state which is conformal at high
temperatures but introduces a mass scale, denoted by $m$, that parameterises the
departure from conformality.  This approach takes inspiration from
parameterisations of the QCD equation of state proposed to emulate the results
of lattice computations (see e.g. Ref.~\cite{HotQCD:2014kol}), but is simple
enough to allow some analytic calculations. We will also neglect bulk viscosity,
which is known to be very small at high temperature, rising sharply only close
to the transition temperature. This simplification will be taken a step further
by neglecting the bulk pressure entirely, which leads to the technical
convenience that the extended phase space remains $3$-dimensional so it is very
easy to compare with conformal MIS theory.  At temperatures above the chiral
transition, the resulting model captures some of the physics at a qualitative
level.

The model equation of state we adopt reads:
\begin{equation}
  \label{eq:eos}
  p = \f{a}{3} \f{T^4}{1 + \f{m^2}{T^2}}\,,
\end{equation}
where $a, m$ are constant parameters. When $T\gg m$, at leading order one recovers
the conformal result $p\sim a T^4/3$, with the subleading correction of a form
appropriate for a gas of massive particles.
Standard thermodynamic relations imply that the
entropy density $s$ and energy density $\ed$ are given by
\begin{subequations}
  \begin{align}
    \label{eq:thermo}
    s &= \f{4 a}{3} T^3 \f{\left( 1 + \f{3m^2}{2T^2} \right)}{\left( 1 +
    \f{m^2}{T^2} \right)^2}\,,
    \\
    \ed &= a T^4 \f{\left( 1 + \f{5m^2}{3T^2} \right)}{\left( 1 + \f{m^2}{T^2}
    \right)^2}\,.
  \end{align}
\end{subequations}
The energy momentum tensor in this case acquires a non-zero trace.
The mass scale $m$ happens to coincide with the position of the maximum of the
normalised trace anomaly
\begin{equation}
  \label{eq:tran}
  \Delta(T) \equiv \frac{\ed - 3 p}{T^4} = \frac{2 a}{3} \f{m^2 T^2}{
    \left(m^2+T^2\right)^2}\,.
\end{equation}
In the following, when numerical values are needed, we
set $m=200$ MeV and take $a = 8 \pi^2/15$, which is the value appropriate for
pure \(SU(3)\) Yang-Mills theory in the limit of high temperature.
The speed of sound is
\begin{equation}
  \label{eq:speed}
  c_s^2(T) = \f{1}{3} \frac{\left(1 + \frac{m^2}{T^2}\right) \left(1+\frac{3
        m^2}{2 T^2}\right)}{\left(1 + \frac{17 m^2}{6 T^2}+ \frac{5 m^4}{2
        T^4}\right)} \,.
\end{equation}
Its asymptotic behaviour at $T\gg m$ and $T\ll m$ is
\begin{subequations}
  \begin{align}
    c_s^2(T) &\sim \frac{1}{3} - \frac{1}{9} \frac{m^2}{T^2}
    + O\left(\frac{m^4}{T^4}\right)\,,
    \\
    \qquad
    c_s^2(T) &\sim\frac{1}{5} + \f{8}{75}\frac{T^2}{m^2}
    + O\left(\frac{T^4}{m^4}\right)\,.
  \end{align}
\end{subequations}

To study boost invariant flow with the nonconformal equation of state given in
\cref{eq:eos} it is useful to start with the ideal fluid. Imposing the
symmetries of Bjorken flow, conservation of the energy momentum tensor implies
\begin{equation}
  \label{eq:ideal}
  \tau \p_\tau \ln T + c_s(T) = 0\,.
\end{equation}
Unlike the ideal fluid evolution equation for the conformal case,
this equation does not have an exact power law
solution, but it can be directly integrated. At late time
\begin{equation}
  \label{eq:asymTideal}
  T \sim  \frac{\Lambda}{(\Lambda\tau)^{1/5}}
  + \frac{4 \Lambda^2}{15 m^2}\frac{\Lambda}{(\Lambda\tau)^{3/5}}
  + \dots\,,
\end{equation}
where $\Lambda$ is an integration constant.  The leading asymptotic behaviour
reflects the limiting value of the speed of sound at low temperature,
\cref{eq:speed}.
All powers of the proper time appear in the above series, but the entropy
density falls off as $1/\tau$ exactly, which expresses the conservation of
entropy in ideal fluid flow.

We now turn to the MIS evolution equations for the model defined by
\cref{eq:eos}. For simplicity, we will keep the transport coefficients the same
as in the conformal case, given in \cref{eq:conformal}.  Proceeding in this way
one obtains
\begin{subequations}
  \begin{align}
    \tau \p_\tau\log T  &= - \frac{1}{3} R_1 + \frac{1}{18} R_2\, \pa \,,
  \label{eq:ncmisbjT}
    \\
    \tau \p_\tau \pa  &=
    S_1\, \frac{8 C_\eta }{C_\pi } - S_2\, \frac{\tau  T }{C_\pi } \pa
    -\frac{2}{9} S_3\, \pa^2\,,
  \label{eq:ncmisbjA}
  \end{align}
\end{subequations}
where the $R_k, S_k$ are rational functions of the effective temperature:
\begin{subequations}
  \label{eq:pfuncs}
  \begin{align}
    R_1 &= \frac{6 \left(m^2+T^2\right)^2}{15 m^4+17 m^2 T^2+6 T^4}\,,\\
    R_2 &= 1-\frac{2 m^2 \left(3 m^2+T^2\right)}{15 m^4+17 m^2 T^2+6 T^4}\,,\\
    S_1 &= \frac{3 m^2+2 T^2}{2 \left(m^2+T^2\right)} \,,\\
    S_2 &= 1 + \frac{2 m^2 (3 m^2-2 T^2)C_\pi }{45 m^4 \tau
    T+51 m^2 \tau  T^3+18 \tau T^5}\,,\\
    S_3 &= 1-\frac{2 m^2 \left(3 m^2+T^2\right)}{15 m^4+17 m^2 T^2+6 T^4}\,.
  \end{align}
\end{subequations}
They all equal unity when $m$ is set to zero -- in that case \cref{eq:ncmisbjT,eq:ncmisbjA}
reduce to \cref{eq:misbjT,eq:misbjA}.
Importantly, this system of equations is singular at $\tau=0$.

\begin{figure}[t]
  \centering
  \includegraphics[width=0.45\linewidth]{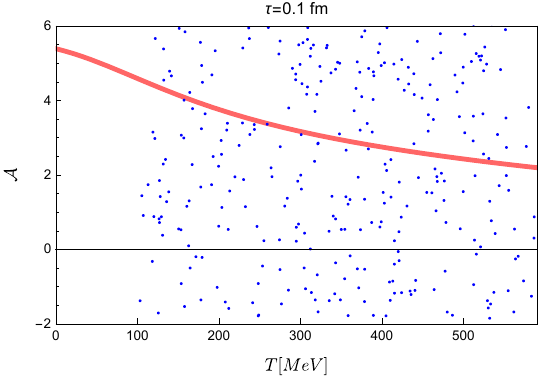}
  \includegraphics[width=0.45\linewidth]{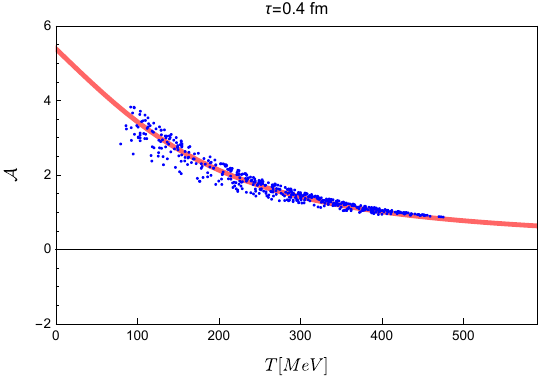}
  \includegraphics[width=0.45\linewidth]{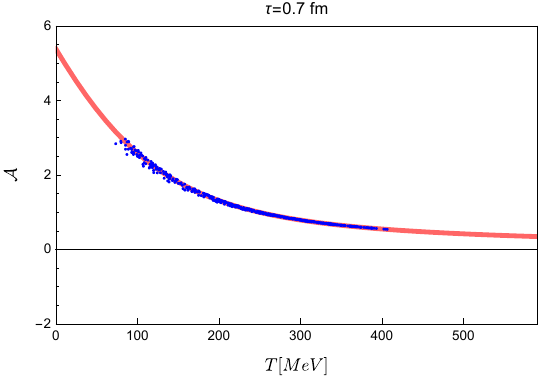}
  \includegraphics[width=0.45\linewidth]{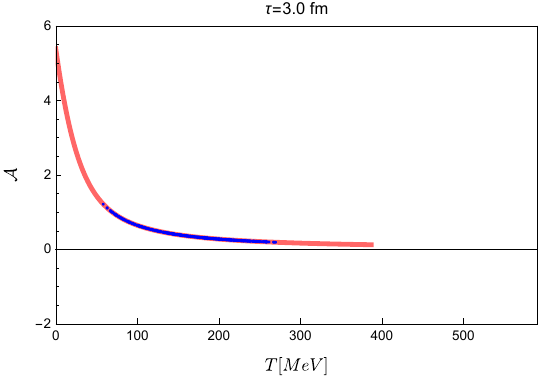}
  \caption{The evolution of the attractor section (the red curve) on a set of constant
    proper time slices in the extended phase space of the MIS theory defined 
    by \cref{eq:ncmisbjT,eq:ncmisbjA}. At the initial time ($0.1$ fm)
    the initial conditions for the solutions plotted as blue dots are set uniformly
    in a range of initial temperatures and pressure anisotropies. At $\tau=0.5$ fm
    all the solutions are in the vicinity of the attractor, with pressure
    anisotropies ranging from around $0.5$ to about $3.5$. Subsequent evolution follows
    the (evolving) attractor. The range of temperatures and pressure anisotropies
    becomes narrower as time goes on, tending toward zero as the system cools.
    The values of the transport coefficients used to create these plots were the
  same as in the conformal case presented in \cref{sec:confomis}.}
  \label{fig:ncattr}
\end{figure}

As in the conformal case, \cref{eq:ncmisbjT,eq:ncmisbjA} imply a second order differential
equation for $T$ alone (or $\pa$ alone), which can be convenient but not
essential for the derivation of asymptotic solutions.
At late times, $\tau\to\infty$, one finds the asymptotic behaviour
\begin{subequations}
  \label{eq:ncasym}
  \begin{align}
    T &\sim \f{\Lambda}{(\Lambda\tau)^{1/5}}\left(
      1 + \frac{4 \Lambda ^2}{15 m^2} \f{1}{(\Lambda\tau)^{2/5}} \right.
    + \left. \left(\frac{\Lambda ^4}{45 m^4}-\frac{C_\eta }{3}
\right)\f{1}{(\Lambda\tau)^{4/5}}+\dots \right) \,,
    \label{eq:ncasT}
    \\
    \pa &\sim \f{12 C_\eta}{(\Lambda\tau)^{4/5}}
    \left(
      1 - \f{3 \Lambda^2}{5 m^2} \f{1}{(\Lambda\tau)^{2/5}}\right.
    + \left. \f{1}{3}\left(C_\eta + 2 C_\pi + \f{22 \Lambda^4}{15 m^4}\right)
\f{1}{(\Lambda\tau)^{4/5}} + \dots \right) \,.
    \label{eq:ncasA}
  \end{align}
\end{subequations}
The dependence on the relaxation time enters at one order higher than in
the conformal case due to the subleading ideal fluid contribution in
\cref{eq:asymTideal}. Only one integration constant, $\Lambda$, appears in this
series. The other integration constant, denoted below by $\sigma$, enters as an
exponentially-suppressed transseries correction to the above power-law
asymptotics:
\begin{subequations}
    \label{eq:ts}
  \begin{align}
    T &\sim \f{\Lambda}{(\Lambda\tau)^{1/5}}\Phi_0(\Lambda\tau, \Lambda)
    + \sigma \Phi_1(\Lambda\tau, \Lambda) (\Lambda \tau)^\beta
    e^{
      -\frac{5 }{4 C_\pi} (\Lambda \tau)^{4/5}
      - \f{2 \Lambda^2}{3 C_\pi m^2} (\Lambda\tau)^{2/5}
    }
    + \dots\\
    \pa &\sim \f{12 C_\eta}{(\Lambda\tau)^{4/5}}
    \Psi_0(\Lambda\tau, \Lambda)
    + \tilde{\sigma} \Psi_1(\Lambda\tau, \Lambda) (\Lambda \tau)^\beta
     e^{ -\frac{5 }{4 C_\pi} (\Lambda \tau)^{4/5}
      - \f{2 \Lambda^2}{3 C_\pi m^2} (\Lambda\tau)^{2/5}}
    + \dots
  \end{align}
\end{subequations}
where
\begin{equation}
  \beta \equiv \frac{17}{3} - \frac{5 C_\eta }{3 C_\pi }+
  \frac{1}{9 C_\pi} \frac{\Lambda ^4}{m^4}
\end{equation}
and $\Phi_k(\Lambda\tau, \Lambda)$, $\Psi_k(\Lambda\tau, \Lambda)$ are power
series in $\Lambda\tau$ with coefficients that depend on $\Lambda$. This is in
contrast to the case of conformal MIS theory, where the corresponding
coefficients appearing in \cref{eq:mists.a,eq:mists.b} are universal for all solutions. The
constant $\sigma$ is the transseries parameter and $\tilde{\sigma}$ can be expressed in
terms of $\sigma$. The leading terms in $\Phi_0(\Lambda\tau, \Lambda)$ and
$\Psi_0(\Lambda\tau, \Lambda)$ are given in \cref{eq:ncasT,eq:ncasA}; coefficients of
the remaining series can be calculated using the equations of motion.
While this
transseries has some novel elements as compared to conformal MIS, the basic
picture of hydrodynamisation does not differ from that discussed in
\cref{sec:confomis}. It follows the pattern familiar from earlier
studies~\cite{Heller:2015dha,Basar:2015ava,Aniceto:2015mto,Aniceto:2022dnm,An:2023yfq}:
a part of the data contained in the initial conditions enters via
exponentially-suppressed terms which are nonperturbative from the point of view
of classical asymptotics, that is, which vanish faster than any power of the
proper time.

At late times, \cref{eq:ncasym} can be interpreted as a parametric
representation of the section of the attractor on a slice of fixed $\tau$, with
$\Lambda$ as the parameter along this curve. Clearly, the result of eliminating
this parameter will depend on the proper time slice, so the attractor sections
evolve in time. This argument can also be applied to earlier times by using the
transseries given in \cref{eq:ts}. To assess whether the attractor extends into
the far from equilibrium domain, we proceed numerically as in the previous
sections. The asymptotic behaviour of solutions at $\tau= 0$ leads to the
conclusion that solutions for which the pressure anisotropy remains finite
satisfy the same initial conditions as in the conformal case, $\pa\sim\pa_+$,
where $\pa_+$ is given in \cref{eq:attra}. This way one obtains the set of plots
shown in \cref{fig:ncattr}.
Both here and in \cref{fig:attra} we have used the $(T, \pa, \tau)$
parameterisation of the extended phase space so one can easily compare the
evolution of the attractor section
to its
behaviour in conformal MIS theory. 
The shape of the attractor here is different,
but the role it plays in the dynamics of the system is the
same: generic solutions first coalesce on the attractor and then evolve along
it. In this model there is no parameterisation in
which the attractor sections would be the same on each slice. Despite this,
there is no doubt when comparing \cref{fig:ncattr} and \cref{fig:attra} that
generic solutions follow the attractor in a very similar way.  While these results
cannot be taken as indication of what will be found in
systems with phase spaces of higher dimension, such as those of
Refs.~\cite{Jaiswal:2014isa,Denicol:2014mca,Denicol:2014vaa,Chen:2021wwh,Chattopadhyay:2021ive,Jaiswal:2022udf,Jaiswal:2021uvv},
the same approach can be applied.

\section{Summary and outlook}
\label{sec:outlook}


The attractor of conformal MIS theory was originally defined as a specific solution
to a first order ordinary differential equation satisfied by the pressure
anisotropy $\pa(w)$. The partial decoupling of evolution equations which is
crucial for such an attractor solution to exist is usually not attainable, so in
order to address more general situations
it is useful to adopt a different perspective, recognising
the attractor as a hypersurface in an extended phase space. Two different
approaches to determining this manifold have been proposed:
Ref.~\cite{Heller:2020anv} introduced a local notion of attractor that is not
connected to any asymptotic limits, while in Ref.~\cite{An:2023yfq} the
attractor hypersurface was defined by imposing regularity of the pressure
anisotropy at asymptotically small proper time. These two approaches should be
compatible, and we have illustrated this in two cases where an attractor
solution is known to exist: conformal MIS theory and the Denicol-Noronha
model. We have demonstrated that in these cases, in the appropriate coordinates,
the projection of the attractor hypersurface obtained using regularity coincides
with the attractor solution in the sense of Ref.~\cite{Heller:2015dha}, and it
does so at all times. This we showed formally by reinterpreting the transseries
representation of the attractor, as well as numerically by observing that
attractor sections in these cases are independent of the proper time and thus
can be projected.

We took these developments a step further by proposing that in situations where
the evolution equations are singular at early time, regularity conditions should
uniquely determine the attractor hypersurface regardless of whether the
evolution equations can be partially decoupled or not. To illustrate the
efficacy of this prescription, we have considered a model of MIS theory where
the evolution equations do not decouple, but regularity at early time still
determines a unique attractor hypersurface and generic solutions evolve toward
it already at high pressure anisotropy: the essential feature which leads to
early-time attraction is the singularity at $\tau=0$. Importantly, this approach
works directly with the equations describing evolution in proper time, without
requiring any special coordinates. The basic picture of hydrodynamisation
remains the same: it begins with a rapid approach to the attractor followed by
exponential, nonhydrodynamic mode decay (captured by a transseries containing
contributions carrying nonhydrodynamic data present in the initial conditions)
and then finally a hydrodynamic approach to equilibrium~\cite{Heller:2020anv}.

There are several possible applications of the approach advocated
here. Perhaps the simplest one would be to reconsider the fluid-dynamical
description of the massive relativistic
gas~\cite{Jaiswal:2014isa,Denicol:2014mca,Denicol:2014vaa,Chen:2021wwh} from
this perspective. A little further afield is the treatment of nonconformal
kinetic theory
models~\cite{Jaiswal:2022udf,Chattopadhyay:2021ive,Jaiswal:2021uvv}.  It may
also be interesting to consider the impact of a possible singularity at early
time on the hydrodynamic attractors that were recently described in the context
of a Fermi gas close to unitarity with a time-dependent scattering
length~\cite{Fujii:2024yce,Mazeliauskas:2025jyi,Heller:2025yxm}. In this
setting, early-time attraction may appear because the system is externally
driven in a specific manner. This drive could lead to consequences similar to
the boost invariance of particle production in QCD and could result in a
singularity of the equations of motion at early time.

From a broader perspective, universal features in the dynamics of nonequilibrium
systems can arise in various ways. The appearance of attractors in
boost-invariant flow is linked to the strong longitudinal expansion and the
ensuing singularity at early proper time, resulting in the emergence of
prehydrodynamic universality in the sense of many different initial states being
dynamically driven to a specific region in phase space. There are also other
examples of universal far from equilibrium behaviour, such as nonthermal fixed
points~\cite{Berges:2008wm,Berges:2013fga,Mazeliauskas:2018yef,Berges:2020fwq},
which arise for special classes of initial conditions.  It would be of great
interest to understand how these phenomena relate to attractors of the type
discussed here, particularly in the context of heavy ion collisions.


\acknowledgments{
I would like to thank Jean-Paul Blaizot, Gabriel Denicol, Tuomas Lappi and
Derek Teaney for helpful conversations.  This research was supported by the
National Science Centre, Poland, under Grant No. 2021/41/B/ST2/02909.
}

\bibliographystyle{JHEP}

\bibliography{global}

\providecommand{\href}[2]{#2}\begingroup\raggedright\begin{thebibliography}{10}

\bibitem{Muller:1967zza}
I.~Muller, \emph{{Zum Paradoxon der Warmeleitungstheorie}},
  \href{https://doi.org/10.1007/BF01326412}{\emph{Z. Phys.} {\bfseries 198}
  (1967) 329}.

\bibitem{Israel:1976tn}
W.~Israel, \emph{{Nonstationary irreversible thermodynamics: A Causal
  relativistic theory}},
  \href{https://doi.org/10.1016/0003-4916(76)90064-6}{\emph{Annals Phys.}
  {\bfseries 100} (1976) 310}.

\bibitem{Israel:1979wp}
W.~Israel and J.M.~Stewart, \emph{{Transient relativistic thermodynamics and
  kinetic theory}},
  \href{https://doi.org/10.1016/0003-4916(79)90130-1}{\emph{Annals Phys.}
  {\bfseries 118} (1979) 341}.

\bibitem{Schenke:2021mxx}
B.~Schenke, \emph{{The smallest fluid on Earth}},
  \href{https://doi.org/10.1088/1361-6633/ac14c9}{\emph{Rept. Prog. Phys.}
  {\bfseries 84} (2021) 082301}
  [\href{https://arxiv.org/abs/2102.11189}{{\ttfamily 2102.11189}}].

\bibitem{Heller:2015dha}
M.P.~Heller and M.~Spaliński, \emph{{Hydrodynamics Beyond the Gradient
  Expansion: Resurgence and Resummation}},
  \href{https://doi.org/10.1103/PhysRevLett.115.072501}{\emph{Phys. Rev. Lett.}
  {\bfseries 115} (2015) 072501}
  [\href{https://arxiv.org/abs/1503.07514}{{\ttfamily 1503.07514}}].

\bibitem{Blaizot:2017ucy}
J.-P.~Blaizot and L.~Yan, \emph{{Fluid dynamics of out of equilibrium boost
  invariant plasmas}},
  \href{https://doi.org/10.1016/j.physletb.2018.02.058}{\emph{Phys. Lett. B}
  {\bfseries 780} (2018) 283}
  [\href{https://arxiv.org/abs/1712.03856}{{\ttfamily 1712.03856}}].

\bibitem{Blaizot:2019scw}
J.-P.~Blaizot and L.~Yan, \emph{{Emergence of hydrodynamical behavior in
  expanding ultra-relativistic plasmas}},
  \href{https://doi.org/10.1016/j.aop.2019.167993}{\emph{Annals Phys.}
  {\bfseries 412} (2020) 167993}
  [\href{https://arxiv.org/abs/1904.08677}{{\ttfamily 1904.08677}}].

\bibitem{Bjorken:1982qr}
J.D.~Bjorken, \emph{{Highly Relativistic Nucleus-Nucleus Collisions: The
  Central Rapidity Region}},
  \href{https://doi.org/10.1103/PhysRevD.27.140}{\emph{Phys. Rev. D} {\bfseries
  27} (1983) 140}.

\bibitem{Carrington:2020ssh}
M.E.~Carrington, A.~Czajka and S.~Mrowczynski, \emph{{The energy-momentum
  tensor at the earliest stage of relativistic heavy-ion collisions}},
  \href{https://doi.org/10.1140/epja/s10050-021-00600-x}{\emph{Eur. Phys. J. A}
  {\bfseries 58} (2022) 5} [\href{https://arxiv.org/abs/2012.03042}{{\ttfamily
  2012.03042}}].

\bibitem{Carrington:2021qvi}
M.E.~Carrington, A.~Czajka and S.~Mr{\'o}wczy{\'n}ski, \emph{{Physical
  characteristics of glasma from the earliest stage of relativistic heavy ion
  collisions}}, \href{https://doi.org/10.1103/PhysRevC.106.034904}{\emph{Phys.
  Rev. C} {\bfseries 106} (2022) 034904}
  [\href{https://arxiv.org/abs/2105.05327}{{\ttfamily 2105.05327}}].

\bibitem{Carrington:2024utf}
M.E.~Carrington, S.~Mrowczynski and J.-Y.~Ollitrault, \emph{{Hydrodynamic-like
  behavior of glasma}},
  \href{https://doi.org/10.1103/PhysRevC.110.054903}{\emph{Phys. Rev. C}
  {\bfseries 110} (2024) 054903}
  [\href{https://arxiv.org/abs/2406.14463}{{\ttfamily 2406.14463}}].

\bibitem{Chen:2024pez}
S.~Chen and S.~Shi, \emph{{Attractor for (1+1)D viscous hydrodynamics with
  general rapidity distribution}},
  \href{https://doi.org/10.1103/PhysRevC.111.L021902}{\emph{Phys. Rev. C}
  {\bfseries 111} (2025) L021902}
  [\href{https://arxiv.org/abs/2407.15209}{{\ttfamily 2407.15209}}].

\bibitem{Nugara:2023eku}
V.~Nugara, S.~Plumari, L.~Oliva and V.~Greco, \emph{{Far-from-equilibrium
  attractors with full relativistic Boltzmann approach in boost-invariant and
  non-boost-invariant systems}},
  \href{https://doi.org/10.1140/epjc/s10052-024-13227-1}{\emph{Eur. Phys. J. C}
  {\bfseries 84} (2024) 861}
  [\href{https://arxiv.org/abs/2311.11921}{{\ttfamily 2311.11921}}].

\bibitem{Kurkela:2019set}
A.~Kurkela, W.~van~der Schee, U.A.~Wiedemann and B.~Wu, \emph{{Early- and
  Late-Time Behavior of Attractors in Heavy-Ion Collisions}},
  \href{https://doi.org/10.1103/PhysRevLett.124.102301}{\emph{Phys. Rev. Lett.}
  {\bfseries 124} (2020) 102301}
  [\href{https://arxiv.org/abs/1907.08101}{{\ttfamily 1907.08101}}].

\bibitem{Ambrus:2021sjg}
V.E.~Ambrus, S.~Busuioc, J.A.~Fotakis, K.~Gallmeister and C.~Greiner,
  \emph{{Bjorken flow attractors with transverse dynamics}},
  \href{https://doi.org/10.1103/PhysRevD.104.094022}{\emph{Phys. Rev. D}
  {\bfseries 104} (2021) 094022}
  [\href{https://arxiv.org/abs/2102.11785}{{\ttfamily 2102.11785}}].

\bibitem{Ambrus:2021fej}
V.E.~Ambrus, S.~Schlichting and C.~Werthmann, \emph{{Development of transverse
  flow at small and large opacities in conformal kinetic theory}},
  \href{https://doi.org/10.1103/PhysRevD.105.014031}{\emph{Phys. Rev. D}
  {\bfseries 105} (2022) 014031}
  [\href{https://arxiv.org/abs/2109.03290}{{\ttfamily 2109.03290}}].

\bibitem{Ambrus:2022koq}
V.E.~Ambrus, S.~Schlichting and C.~Werthmann, \emph{{Opacity dependence of
  transverse flow, preequilibrium, and applicability of hydrodynamics in
  heavy-ion collisions}},
  \href{https://doi.org/10.1103/PhysRevD.107.094013}{\emph{Phys. Rev. D}
  {\bfseries 107} (2023) 094013}
  [\href{https://arxiv.org/abs/2211.14379}{{\ttfamily 2211.14379}}].

\bibitem{Nugara:2024net}
V.~Nugara, V.~Greco and S.~Plumari, \emph{{Far-from-equilibrium attractors with
  Full Relativistic Boltzmann approach in 3+1D: moments of distribution
  function and~anisotropic flows $v_n$}},
  \href{https://doi.org/10.1140/epjc/s10052-025-14029-9}{\emph{Eur. Phys. J. C}
  {\bfseries 85} (2025) 311}
  [\href{https://arxiv.org/abs/2409.12123}{{\ttfamily 2409.12123}}].

\bibitem{An:2023yfq}
X.~An and M.~Spali\'nski, \emph{{QGP physics from attractor perturbations}},
  \href{https://doi.org/10.1103/PhysRevD.110.114043}{\emph{Phys. Rev. D}
  {\bfseries 110} (2024) 114043}
  [\href{https://arxiv.org/abs/2312.17237}{{\ttfamily 2312.17237}}].

\bibitem{Heller:2020anv}
M.P.~Heller, R.~Jefferson, M.~Spali\'nski and V.~Svensson, \emph{{Hydrodynamic
  Attractors in Phase Space}},
  \href{https://doi.org/10.1103/PhysRevLett.125.132301}{\emph{Phys. Rev. Lett.}
  {\bfseries 125} (2020) 132301}
  [\href{https://arxiv.org/abs/2003.07368}{{\ttfamily 2003.07368}}].

\bibitem{Spalinski:2022cgj}
M.~Spali\'nski, \emph{{Initial State and Approach to Equilibrium}},
  \href{https://doi.org/10.5506/APhysPolBSupp.16.1-A9}{\emph{Acta Phys. Polon.
  Supp.} {\bfseries 16} (2023) 9}
  [\href{https://arxiv.org/abs/2209.13849}{{\ttfamily 2209.13849}}].

\bibitem{Jankowski:2023fdz}
J.~Jankowski and M.~Spali\'nski, \emph{{Hydrodynamic attractors in
  ultrarelativistic nuclear collisions}},
  \href{https://doi.org/10.1016/j.ppnp.2023.104048}{\emph{Prog. Part. Nucl.
  Phys.} {\bfseries 132} (2023) 104048}
  [\href{https://arxiv.org/abs/2303.09414}{{\ttfamily 2303.09414}}].

\bibitem{Wiggins}
S.~Wiggins, \emph{Introduction to Applied Nonlinear Dynamical Systems and
  Chaos}, Springer, 2nd~ed. (1990).

\bibitem{Florkowski:2017olj}
W.~Florkowski, M.P.~Heller and M.~Spali{\'n}ski, \emph{{New theories of
  relativistic hydrodynamics in the LHC era}},
  \href{https://doi.org/10.1088/1361-6633/aaa091}{\emph{Rept. Prog. Phys.}
  {\bfseries 81} (2018) 046001}
  [\href{https://arxiv.org/abs/1707.02282}{{\ttfamily 1707.02282}}].

\bibitem{Soloviev:2021lhs}
A.~Soloviev, \emph{{Hydrodynamic attractors in heavy ion collisions: a
  review}}, \href{https://doi.org/10.1140/epjc/s10052-022-10282-4}{\emph{Eur.
  Phys. J. C} {\bfseries 82} (2022) 319}
  [\href{https://arxiv.org/abs/2109.15081}{{\ttfamily 2109.15081}}].

\bibitem{Aniceto:2022dnm}
I.~Aniceto, D.~Hasenbichler and A.O.~Daalhuis, \emph{{The late to early time
  behaviour of an expanding plasma: hydrodynamisation from exponential
  asymptotics}}, \href{https://doi.org/10.1088/1751-8121/acc61d}{\emph{J. Phys.
  A} {\bfseries 56} (2023) 195201}
  [\href{https://arxiv.org/abs/2207.02868}{{\ttfamily 2207.02868}}].

\bibitem{zwillinger}
D.~Zwillinger, \emph{CRC Standard Mathematical Tables and Formulas}, CRC
  (2018).

\bibitem{Denicol:2017lxn}
G.S.~Denicol and J.~Noronha, \emph{{Analytical attractor and the divergence of
  the slow-roll expansion in relativistic hydrodynamics}},
  \href{https://doi.org/10.1103/PhysRevD.97.056021}{\emph{Phys. Rev. D}
  {\bfseries 97} (2018) 056021}
  [\href{https://arxiv.org/abs/1711.01657}{{\ttfamily 1711.01657}}].

\bibitem{HotQCD:2014kol}
{\scshape HotQCD} collaboration, \emph{{Equation of state in ( 2+1 )-flavor
  QCD}}, \href{https://doi.org/10.1103/PhysRevD.90.094503}{\emph{Phys. Rev. D}
  {\bfseries 90} (2014) 094503}
  [\href{https://arxiv.org/abs/1407.6387}{{\ttfamily 1407.6387}}].

\bibitem{Basar:2015ava}
G.~Basar and G.V.~Dunne, \emph{{Hydrodynamics, resurgence, and
  transasymptotics}},
  \href{https://doi.org/10.1103/PhysRevD.92.125011}{\emph{Phys. Rev. D}
  {\bfseries 92} (2015) 125011}
  [\href{https://arxiv.org/abs/1509.05046}{{\ttfamily 1509.05046}}].

\bibitem{Aniceto:2015mto}
I.~Aniceto and M.~Spali\'nski, \emph{{Resurgence in Extended Hydrodynamics}},
  \href{https://doi.org/10.1103/PhysRevD.93.085008}{\emph{Phys. Rev. D}
  {\bfseries 93} (2016) 085008}
  [\href{https://arxiv.org/abs/1511.06358}{{\ttfamily 1511.06358}}].

\bibitem{Jaiswal:2014isa}
A.~Jaiswal, R.~Ryblewski and M.~Strickland, \emph{{Transport coefficients for
  bulk viscous evolution in the relaxation time approximation}},
  \href{https://doi.org/10.1103/PhysRevC.90.044908}{\emph{Phys. Rev. C}
  {\bfseries 90} (2014) 044908}
  [\href{https://arxiv.org/abs/1407.7231}{{\ttfamily 1407.7231}}].

\bibitem{Denicol:2014mca}
G.S.~Denicol, W.~Florkowski, R.~Ryblewski and M.~Strickland, \emph{{Shear-bulk
  coupling in nonconformal hydrodynamics}},
  \href{https://doi.org/10.1103/PhysRevC.90.044905}{\emph{Phys. Rev. C}
  {\bfseries 90} (2014) 044905}
  [\href{https://arxiv.org/abs/1407.4767}{{\ttfamily 1407.4767}}].

\bibitem{Denicol:2014vaa}
G.S.~Denicol, S.~Jeon and C.~Gale, \emph{{Transport Coefficients of Bulk
  Viscous Pressure in the 14-moment approximation}},
  \href{https://doi.org/10.1103/PhysRevC.90.024912}{\emph{Phys. Rev. C}
  {\bfseries 90} (2014) 024912}
  [\href{https://arxiv.org/abs/1403.0962}{{\ttfamily 1403.0962}}].

\bibitem{Chen:2021wwh}
Z.~Chen and L.~Yan, \emph{{Hydrodynamic attractor in the nonconformal Bjorken
  flow}}, \href{https://doi.org/10.1103/PhysRevC.105.024910}{\emph{Phys. Rev.
  C} {\bfseries 105} (2022) 024910}
  [\href{https://arxiv.org/abs/2109.06658}{{\ttfamily 2109.06658}}].

\bibitem{Chattopadhyay:2021ive}
C.~Chattopadhyay, S.~Jaiswal, L.~Du, U.~Heinz and S.~Pal, \emph{{Non-conformal
  attractor in boost-invariant plasmas}},
  \href{https://doi.org/10.1016/j.physletb.2021.136820}{\emph{Phys. Lett. B}
  {\bfseries 824} (2022) 136820}
  [\href{https://arxiv.org/abs/2107.05500}{{\ttfamily 2107.05500}}].

\bibitem{Jaiswal:2022udf}
S.~Jaiswal, J.-P.~Blaizot, R.S.~Bhalerao, Z.~Chen, A.~Jaiswal and L.~Yan,
  \emph{{From moments of the distribution function to hydrodynamics: The
  nonconformal case}},
  \href{https://doi.org/10.1103/PhysRevC.106.044912}{\emph{Phys. Rev. C}
  {\bfseries 106} (2022) 044912}
  [\href{https://arxiv.org/abs/2208.02750}{{\ttfamily 2208.02750}}].

\bibitem{Jaiswal:2021uvv}
S.~Jaiswal, C.~Chattopadhyay, L.~Du, U.~Heinz and S.~Pal, \emph{{Nonconformal
  kinetic theory and hydrodynamics for Bjorken flow}},
  \href{https://doi.org/10.1103/PhysRevC.105.024911}{\emph{Phys. Rev. C}
  {\bfseries 105} (2022) 024911}
  [\href{https://arxiv.org/abs/2107.10248}{{\ttfamily 2107.10248}}].

\bibitem{Fujii:2024yce}
K.~Fujii and T.~Enss, \emph{{Hydrodynamic Attractor in Ultracold Atoms}},
  \href{https://doi.org/10.1103/PhysRevLett.133.173402}{\emph{Phys. Rev. Lett.}
  {\bfseries 133} (2024) 173402}
  [\href{https://arxiv.org/abs/2404.12921}{{\ttfamily 2404.12921}}].

\bibitem{Mazeliauskas:2025jyi}
A.~Mazeliauskas and T.~Enss, \emph{{Hydrodynamic attractor in periodically
  driven ultracold quantum gases}},
  \href{https://arxiv.org/abs/2501.19240}{{\ttfamily 2501.19240}}.

\bibitem{Heller:2025yxm}
M.P.~Heller and C.~Werthmann, \emph{{Early time hydrodynamic attractor in a
  nearly-unitary Fermi gas}},
  \href{https://arxiv.org/abs/2507.02838}{{\ttfamily 2507.02838}}.

\bibitem{Berges:2008wm}
J.~Berges, A.~Rothkopf and J.~Schmidt, \emph{{Non-thermal fixed points:
  Effective weak-coupling for strongly correlated systems far from
  equilibrium}},
  \href{https://doi.org/10.1103/PhysRevLett.101.041603}{\emph{Phys. Rev. Lett.}
  {\bfseries 101} (2008) 041603}
  [\href{https://arxiv.org/abs/0803.0131}{{\ttfamily 0803.0131}}].

\bibitem{Berges:2013fga}
J.~Berges, K.~Boguslavski, S.~Schlichting and R.~Venugopalan, \emph{{Universal
  attractor in a highly occupied non-Abelian plasma}},
  \href{https://doi.org/10.1103/PhysRevD.89.114007}{\emph{Phys. Rev. D}
  {\bfseries 89} (2014) 114007}
  [\href{https://arxiv.org/abs/1311.3005}{{\ttfamily 1311.3005}}].

\bibitem{Mazeliauskas:2018yef}
A.~Mazeliauskas and J.~Berges, \emph{{Prescaling and far-from-equilibrium
  hydrodynamics in the quark-gluon plasma}},
  \href{https://doi.org/10.1103/PhysRevLett.122.122301}{\emph{Phys. Rev. Lett.}
  {\bfseries 122} (2019) 122301}
  [\href{https://arxiv.org/abs/1810.10554}{{\ttfamily 1810.10554}}].

\bibitem{Berges:2020fwq}
J.~Berges, M.P.~Heller, A.~Mazeliauskas and R.~Venugopalan, \emph{{QCD
  thermalization: Ab initio approaches and interdisciplinary connections}},
  \href{https://doi.org/10.1103/RevModPhys.93.035003}{\emph{Rev. Mod. Phys.}
  {\bfseries 93} (2021) 035003}
  [\href{https://arxiv.org/abs/2005.12299}{{\ttfamily 2005.12299}}].

\end{thebibliography}\endgroup

\end{document}